\documentclass[aps,onecolumn,showpacs,groupedaddress,nofootinbib]{revtex4}  
\usepackage{graphics,graphicx}  
\usepackage{dcolumn}   
\usepackage{bm}        
\usepackage{amssymb}   
\usepackage{bigints}
\usepackage{xcolor}
\usepackage{tabularx}
\usepackage{ytableau}
\usepackage{tikz}
\usetikzlibrary{calc,arrows.meta}
\usepackage{cancel}
\usepackage{subfigure}
\usepackage{mathtools}

\def\bl{\begin{equation}\begin{aligned}}
\def\el{\end{aligned}\end{equation}}
\def\bpm{\begin{pmatrix}}
\def\epm{\end{pmatrix}}

\def\beal{\begin{aligned}}
\def\eal{\end{aligned}}
\def\be{\begin{equation}}

\def\be{\begin{equation}}
\def\ee{\end{equation}}
\def\bea{\begin{eqnarray}}
\def\eea{\end{eqnarray}}

\begin{document}

\title{The Pentaquark Spectrum from Fermi Statistics}
\author{Luciano Maiani}
\author{Antonio D. Polosa}
\author{Veronica Riquer}
\affiliation{Dipartimento di Fisica and INFN,  Sapienza  Universit\`a di Roma, Piazzale Aldo Moro 2, I-00185 Roma, Italy.}
\email{luciano.maiani@roma1.infn.it}
\email{antonio.polosa@roma1.infn.it}
\email{veronica.riquer@cern.ch}

\date{\today}

\begin{abstract} 
We study compact hidden charm pentaquarks in the Born-Oppenheimer approximation, previously introduced for tetraquarks, assuming the heavy pair to be in a color octet. We show  that Fermi statistics applied to the complex of the three light quarks, also in color octet, requires  $S$-wave pentaquark ground states to consist of {\it three octets of flavour-SU(3)$_f$, two with spin 1/2 and one with spin 3/2}, in line with the observed, strangeness $S=0,-1$, spectrum. Additional lines corresponding to decays into $J/\psi+\Sigma$ and $J/\psi+\Xi$ are predicted. In the language of non-relativistic SU(6), ground state pentaquarks form either a ${\bm{56}}$ or a ${\bm{20}}$ representation, distinguished by presence or absence of pentaquarks decaying in the spin $3/2$ decuplet, e.g. in $J/\psi+\Delta ^{++}$. Observation of a strangeness $S=-2$ or isospin $I=3/2$ pentaquarks would be a clear signature of compact, QCD bound pentaquarks.
\end{abstract}


\maketitle

\section{Introduction}\label{intro}

After the LHCb observation in 2015~\cite{LHCb:2015yax}, hidden charm pentaquark, compact or molecular, have received considerable attention. Three lines are observed for strangeness $S=0$ final state $J/\psi+p$, and two, possibly three, lines for $S=-1$, $J/\psi +\Lambda$~\cite{2019}.

Loosely bound  molecular pentaquark interpretations can be found in~\cite{m1}. Kinematical effects are discussed in~\cite{m2}. An early discussion of compact pentaquarks can be found in~\cite{Maiani:2015vwa,Ali, Lebed:2015tna,4q}.

To describe compact, hidden charm or beauty pentaquarks, the Born-Oppenheimer (BO) approximation is particularly appropriate since it improves when the mass of the heavy constituents increases with respect to the mass of the light ones, see~\cite{noiBO,weinbergQM,pauling}. 
In this framework, one considers  the heavy constituents ($c$ and $\bar c$) as color sources at rest, with fixed relative distance $R$, and computes the lowest energy $E(R)$ of the three light quarks, either analytically, in perturbation theory with techniques borrowed from molecular physics~\cite{noiBO}, or non-perturbatively, in lattice QCD~\cite{Bicudo:2012qt}.  

Besides distance and spin, one has to specify the color quantum numbers of the  sources. For colored quarks, the natural choice is to assume the heavy quark-antiquark pair to be in a color octet, coupled to the light quarks to make an overall color singlet
\be
{\cal P}_{BO}=[{\bar c}({\bm x}_{\bar c})\lambda^A c({\bm x}_c)] B^A\label{newop}
\ee
(sum over $A=1,\dots, 8$ understood). $B^A$ describes the  complex of the three light quarks, which live in a color octet as well.  Specializing to a proton-like pentaquark, we specify $B^A=B^A[u(x_1),u(x_2), d(x)]$ in terms of the light quark coordinates.

In this paper, we address  the restrictions posed by Fermi statistics to the complex of the three light quarks in \eqref{newop}, considering exchange of color, coordinates, flavor and spin. We summarize flavor and spin in representations of non-relativistic SU(6)\,$\supset$\,SU(3)$_f\,\otimes\,$SU(2)$_{\rm spin}$~\cite{Gursey:1992dc}. 

We show that the simplest hypothesis of complete symmetry under coordinates, which would imply pentaquarks in the mixed-symmetry $\bf{70}$ representation of $SU(6)$~\cite{Santopinto:2016pkp}, is excluded. 

The BO approximation  allows mixed symmetry in the coordinates also for ground state, $S$-wave, pentaquarks by distributing quarks in orbitals around the fixed sources. We show that, in this case, one obtains a consistent solution for the SU(6) representations $\bf{56}$ and $\bf{20}$ but not for the $\bf{70}$. 

The elimination of the $\bf{70}$ greatly reduces spin and flavor multiplicity of ground state pentaquarks. 
Including the spin of the $c-\bar c$ pair, we obtain, for either $\bf{56}$ or $\bf{20}$,  {\it three octets of flavour-SU(3)$_f$}, {\it two with spin} 1/2 {\it and one with spin 3/2}, in line with the $S=0,-1$ spectrum observed in~\cite{LHCb:2015yax}.  

Additional lines corresponding to pentaquarks decaying into $J/\psi+\Sigma ~(S=-1)$ and $J/\psi+\Xi ~(S=-2)$ are predicted. 
 
The two alternatives, $\bf{56}$ and $\bf{20}$, are distinguished by the presence or absence of pentaquarks decaying into spin 3/2 resonances, e.g.: ${\cal P}_{(\Delta,S=0)}^{++} \to J/\psi + \Delta^{++}\to J/\psi + p+\pi^+$, which applies to ${\bm {56}}$ only.

The observation of a strangeness $S=-2$ or isospin $I=3/2$ pentaquarks would be clear signatures of compact, QCD bound pentaquarks.

The plan of the paper is as follows. 
We discuss the properties of three light quarks  operators in Sect~\ref{due}.  Sect.~\ref{tre} illustrates the way to deal with the representations of the group of three objects, $S_3$, for color, coordinates, flavor and spin.  In Sect.~\ref{consist} we present the essentials of the BO approximation for pentaquarks and derive a consistency condition on QCD couplings of light to heavy quarks, for quarks distributed in different BO orbitals. In Sect.~\ref{fstat} we illustrate the conditions required by Fermi statistics on the (flavour\,$\otimes$\, spin) wave function of light quarks and in Sect.~\ref{quattro} compute the resulting SU(6) wave functions.  Finally, in Sect.~\ref{result} we presents our results and illustrate the prospects of future calculations of pentaquark mass spectra in the BO approximation. More technical details are contained in four Appendices.

\section{Basic light quark operators}\label{due}
We describe pentaquarks with operators of the generic form given in Eq.~\eqref{newop}
 \be
{\cal P}_{BO}=[{\bar c}({\bm x}_{\bar c})\lambda^A c({\bm x}_c)] B^A \notag
\ee
$B^A$ is obtained from two basic octets, constructed in turn from antisymmetric or symmetric diquark operators. 

Specialising to proton-like tetraquarks 
\begin{enumerate}
\item the antisymmetric octet is
\bea
&& P^A(u_1,d | u_2)=\bar \theta(u_1 d)\lambda^A u_2= u_1^a d^b\, (\epsilon_{abc}\, \lambda^A_{cd})\, u_2^d\notag\\
&&\bar \theta(q q^\prime)_a=\epsilon_{abc }\, q^b q^{\prime c}
\label{antisymm}
\eea
\item the diquark symmetric octet is
\bea
&& \Phi^A(u_1,d | u_2)=\phi^{ad}\, (\epsilon_{abc}\, \lambda^A_{cd})\, u_2^b\notag\\
&& \phi^{ab}(u,d)=(u^a d^b+u^b d^a)
\label{symm}
\eea
 \end{enumerate}
 
We register here a few relevant Fierz transformations
 \bea
T&=&|(\bar Q Q)_{\bm 8} (\bar \theta q)_{\bm 8}\rangle_{\bm 1}=\sqrt{\frac{2}{3}} |(Qq)_{\bar{\bm 3}} (\bar Q \bar \theta)_{\bm 3}\rangle_1-\frac{1}{\sqrt{3}}|(Qq)_{\bm 6} (\bar Q \bar \theta)_{{\bar {\bm 6}}}\rangle_1=\label{tf1}\\
&=&\sqrt{\frac{8}{9}}|(\bar Q q)_{\bm 1}(\bar \theta Q)_{\bm 1}\rangle-\frac{1}{\sqrt{9}}|(\bar Q q)_{\bm 8}(\bar \theta Q)_{\bm 8}\rangle_{\bm 1}\label{tf2}
\eea
where $\bar \theta$ is a generic ${\bar{\bm 3}}$. For a generic $\phi \in {{\bm 6}}$, the Fierz relations read
\bea
&&T=|(\bar Q Q)_{\bm 8} (\phi q)_{\bm 8}\rangle_{\bm 1}=|(\bar Q q)_{\bm 8} (\phi Q)_{\bm 8}\rangle_{\bm 1}=|(Qq)_{\bar {\bm 3}} (\phi\bar Q)_{\bm 3}\rangle_{\bm 1}\label{tff2}
\eea
which follow from the SU(3)$_c$ composition rules: ${{\bm 6}}\otimes {\bm 3}={\bm 8}\oplus{\bm {10}}$ and ${{\bm 6}}\otimes {\bar {\bm 3}}={\bm 3}\oplus{\bm {15}}$. 

Given the form of $B^A$ in terms of $P^A$ and $\Phi^A$, relations \eqref{tf1} to \eqref{tff2} allow  us to find the QCD couplings $qQ$ and $q\Bar Q$.  
We define
\be 
g^2_{cq}=\alpha_s \lambda_{cq}\qquad \lambda_{cq}=\frac{1}{2}(C_2({\bm R})-8/3) \label{gstrong}
\ee
$C_2({\bm R})$ is the quadratic Casimir operator~\footnote{We note the results:
$C_2({\bm 1})=0$; $C_2({\bm R})=C_2({\bar{\bm R}})$; $C_2 ({\bm 3})=4/3$; $C_2({\bm 6})=10/3$;  $C_2({\bm 8})=3$.} of the color representation ${\bm R}$ of the pair $cq$.
 If the pair is in a superposition of two or more SU(3)$_c$ representations with amplitudes $a$, $b$, $\dots$ we use~\cite{noiBO}
\bea
&&T= a\,|(c q)_{{\bm  R}_1}\dots \rangle_{\bm 1}+b\,|(c q)_{{\bm  R}_2}\dots \rangle_{\bm 1}+\dots \nonumber  \\
&&\lambda_{cq }=a^2\lambda_{cq}({\bm R_1})+b^2\lambda_{cq}({\bm R_2})+\dots\label{casimir2}.
\eea

{\bf{\emph{Exchanging quarks.}}}\label{moving} 
Since $P^A$ and $\Phi^A$ are the only octet operators with the given three quarks, we must be able to express $P^A(u_1,d | u_2)$ in terms of octets made by $u_1 u_2$ and $d$, i.e. find $a$ and $b$ such that
\be
P^A(u_1 d | u_2)= a~P^A(u_1 u_2 | d)+ b~\Phi^A(u_1 u_2 | d)\label{superpos}
\ee
Relation~\eqref{superpos} is what we need to move quarks around and bring a given pair together. However, operators $P,~\Phi$ are not equally normalised. Starting from $a$ and $b$, we compute the normalisation factors to obtain a relation between normalised kets
 \be
|(u_1 d)_{{\bar{\bf 3}}},u_2\rangle_{\bm 8}=\alpha |(u_1 u_2)_{{\bar{\bf 3}}}, d\rangle_{\bm 8} + \beta |(u_1 u_2)_{\bf 6}, d \rangle_{\bm 8}
\ee
with $\alpha^2 +\beta^2 =1$.
A simple calculation leads to the transformation table
\begin{table}[htb!]
\centering
   \begin{tabular}{|c|c|c||c|c|}
     \hline
-- & {\footnotesize$|(u_1, u_2)_{\bar{\bf 3}}, d\rangle_{\bf 8}$}&  {\footnotesize$|(u_1, u_2)_{{\bf 6}}, d\rangle_{\bf 8}$}& {\footnotesize$|(u_2, d)_{\bar{\bf 3}}, u_1\rangle_{\bf 8}$}&  {\footnotesize$|(u_2, d)_{{\bf 6}}, u_1\rangle_{\bf 8}$}  \\ \hline
 {\footnotesize$|(u_1, d)_{\bar{\bf 3}}, u_2\rangle_{\bf 8}$}&$+1/2$ & $+\sqrt{3}/2$& $+1/2$ & $-\sqrt{3}/2$\\ \hline
  {\footnotesize$|(u_1, d)_{{\bf 6}}, u_2\rangle_{\bf 8}$}&$\sqrt{3}/2$ & $-1/2$ & $-\sqrt{3}/2$  & $-1/2$\\ \hline
\end{tabular}
 \caption{\footnotesize {Transformation table for quark rearrangements inside $P^A$ and $\Phi^A$ color octets.}}
\label{uno2}
\end{table}

\section{The group $S_3$ and its representations} \label{tre}
The group $S_3$ of the six permutations of three elements can be seen as the group of symmetries of an equilateral triangle by thinking of these as permuting the three vertices.  This group consists of the identiy transformation corresponding to three cycles of length one, two rotations by $120^\circ $ and $240^\circ$, which are permutations corresponding to cycles of length three (e.g. the vertex $a\to b\to c\to a$) and  three reflections in the three altitudes of the triangle, each consisting of a cycle of length two combined with a cycle length one leaving one vertex unchanged. 

The two dimensional representation
~is given by~\cite{georgi}
\be
E=\bpm 1&0\\0&1\\ \epm\qquad D(\sigma_1)=\frac{1}{2}\bpm -1&-\sqrt{3}\\\sqrt{3}&-1\\ \epm \qquad D(\sigma_2)=\frac{1}{2}\bpm -1&\sqrt{3}\\-\sqrt{3}&-1\\ \epm
\ee
and
\be
D(\tau_1)=\bpm -1&0\\0&1\\ \epm\qquad D(\tau_2)=\frac{1}{2}\bpm 1&-\sqrt{3}\\-\sqrt{3}&-1\\ \epm \qquad D(\tau_3)=\frac{1}{2}\bpm 1&\sqrt{3}\\\sqrt{3}&-1\\ \epm
\ee
Having in mind the equilateral triangle
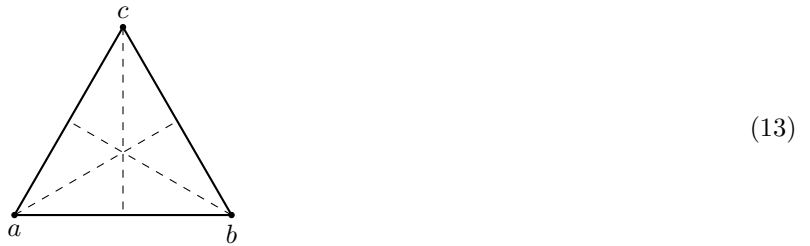
\bl
\begin{tikzpicture}
\coordinate (A) at (0,0);
\coordinate (B) at (1.443,2.5);
\coordinate (C) at (2.886,0);
\draw[thick] (A)--(B)--(C)--cycle;
\draw[dashed] (A)--($(B)!0.5!(C)$);
\draw[dashed] (B)--($(A)!0.5!(C)$);
\draw[dashed] (C)--($(A)!0.5!(B)$);
\filldraw[black] (A) circle (1pt) node[below] {$a$};
\filldraw[black] (B) circle (1pt) node[above] {$c$};
\filldraw[black] (C) circle (1pt) node[below] {$b$};
\end{tikzpicture}
\el
we see that $D(\sigma_1)$ is a anti-clockwise  rotation by $120^\circ$  of the position vectors of the vertices of the triangle (taken from the center), and $D(\sigma_2)$ is a rotation by $-120^\circ$. $D(\tau_1)$ represents a reflection (through the $y$ axis).
A rotation $D(\sigma_1)$ changes $a\mapsto b\mapsto c\mapsto a$. If after this rotation a $D(\tau_1)$ reflection is done, which amounts to $c\rightleftarrows a$, we get $D(\tau_1)D(\sigma_1)=D(\tau_3)$, which corresponds to a $b\rightleftarrows c$ reflection on the original triangle. Similarly $D(\tau_1)D(\sigma_2)=D(\tau_2)$, corresponding to 
a $a\rightleftarrows c$ reflection.

Let us name the eigenvectors of the reflection $D(\tau_1)$ as 
\be
D(\tau_1)M^\lambda =M^\lambda\qquad D(\tau_1)M^\rho =-M^\rho
\ee
Therefore
\bea
&&D(\tau_2)M^\lambda=-\frac{\sqrt{3}}{2}M^\rho-\frac{1}{2}M^\lambda;~~D(\tau_3)M^\lambda=\frac{\sqrt{3}}{2}M^\rho-\frac{1}{2}M^\lambda  
\label{mixref1}\\ 
&&D(\tau_2)M^\rho=\frac{1}{2}M^\rho-\frac{\sqrt{3}}{2}M^\lambda;~~D(\tau_3)M^\rho=\frac{1}{2}M^\rho+\frac{\sqrt{3}}{2}M^\lambda 
\label{mixref2}
\eea

Two different mixed representations of $S_3$ acting on different variables, $M_1$ and $M_2$,  may combine to an $A$, $S$ or $M$ representation, according to the scheme~\cite{roberts}
\bl
S&=\frac{1}{\sqrt{2}}(M_1^\rho M_2^\rho +M_1^\lambda M_2^\lambda)  \quad   &A=\frac{1}{\sqrt{2}}(M_1^\rho M_2^\lambda - M_1^\lambda M_2^\rho)\\
M^\rho&=\frac{1}{\sqrt{2}}(M_1^\rho M_2^\lambda + M_1^\lambda M_2^\rho) \quad  &M^\lambda=\frac{1}{\sqrt{2}}(M_1^\rho M_2^\rho - M_1^\lambda M_2^\lambda) 
\label{composizione}
\el

For color we define
\bea
&& |(q_1,q_2)_{\bar {\bm 3}},q_3\rangle_{\bm 8}= M^\rho \label{asimm}\\
&&|(q_1,q_2)_{{\bm 6}},q_3\rangle_{\bm 8}= M^\lambda \label{simm}
\eea

where $q_1,q_2,q_3$ correspond  to the vertices in the triangle $a,b,c$.

Eqs.~\eqref{mixref1} and  \eqref{mixref2} reproduce the
 results   in Table~\ref{uno2}. The left side columns of Table~\ref{uno2}  correspond to $D(\tau_3)$, or $b\rightleftarrows c$ reflection, whereas the right side columns corresponds to $D(\tau_2)$, or $a\rightleftarrows c$ reflection.

Following \cite{roberts}, we list explicitly in App.~\ref{templates} the  building blocks of mixed and symmetric representations with respect to color, flavour, spin and coordinates, restricting to proton-like states (flavour octet or decuplet).

\section{Born-Oppenheimer approximation: a QCD consistency condition} \label{consist}

We follow the perturbative scheme illustrated in \cite{pauling} for hydrogen molecules and ions in QED and in~\cite{noiBO} for hidden charm hadrons in QCD. The starting point is the interaction of each light particle with the fixed sources, following the instructions given in Sect.~\ref{due}.
 
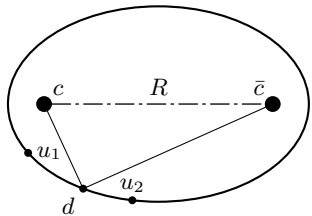
\begin{figure}[h!]
\centering
\begin{tikzpicture}
  \def\a{2} 
  \def\b{1.3} 
  \draw[thick] (0,0) ellipse (\a cm and \b cm); 
  \pgfmathsetmacro{\fd}{sqrt(\a*\a - \b*\b)}
  \coordinate (f1) at (-\fd,0);
  \coordinate (f2) at (\fd,0);
  \fill (f1) circle (3pt);
  \fill (f2) circle (3pt);
  \foreach \i in {210,260,240}{
    \fill ({\a*cos(\i)}, {\b*sin(\i)})  circle (1.5pt);   
  }
  \coordinate (A) at ({\a*cos(210)},{\b*sin(210)});
    \node[right] at (A) {$u_1$};
     \coordinate (B) at ({\a*cos(260)},{\b*sin(260)});
    \node[above] at (B) {$u_2$};
    \coordinate (C) at ({\a*cos(240)},{\b*sin(240)});
    \node[anchor=north east] at (C) {$d$};
    \draw[] (f1)--({\a*cos(240)},{\b*sin(240)})--(f2);
      \node[anchor = south west] at (f1) {$c$};
      \node[anchor=south east] at (f2) {$\bar c$};
     \draw[dotted, line cap=round, line width=0.5pt, dash pattern=on 1pt off 2pt on 7pt off 2pt]  (f1)--(f2);
      \node[above]  at ($(f1)!.5!(f2)$)  {$R$};
\end{tikzpicture}
\caption{\footnotesize{For the three light quarks to be in the same orbital, it is necessary that they carry the same color charge with respect to both  $c$ and $\bar c$.}\label{ellissi}}
\end{figure}
\begin{enumerate}
 \item In the simplest {\it one orbital} scheme, one solves, analitytically or numerically  \cite{pauling}, the Scr\"odinger equation of $q$ in the presence of  the static sources. The orbital is  the corresponding ground state with wave function, $\psi_0(x_q)$ and energy $\epsilon_0$. We represent the orbital in Fig.~\ref{ellissi} with an ellipse around the heavy sources. Similarly to what done in atomic physics, we put the other light quarks in the same orbital, corresponding to the ground state wave function
 \bea
&& \Psi_0(x_{u_d},x_{u_1},x_{u_2})=\psi_0(x_{d})\psi_0(x_{u_1}) \psi_0(x_{u_2}) \notag\\
 && E_0=3\epsilon_0
 \eea
$\Psi_0$ is  symmetric under the exchange of quark coordinates. In the atomic physics language, we attribute occupation number $3$ to the orbital. 

Denoting by $V_{\rm res}$ the sum of the  light-to-light interactions, that did not intervene in the construction of  the orbital, the BO potential, to first order in $V_{res}$, is
\be
V_{\rm BO}=E_0+ \langle \Psi|V_{\rm res}|\Psi\rangle +V_{c\bar c}(R)
\ee
where $V_{c\bar c}(R)$ is the QCD interaction between $c$ and $\bar c$. $V_{BO}$ is obviously a function of $R$ and is the potential of the Schr\"odinger equation of the $c\bar c$ system. 

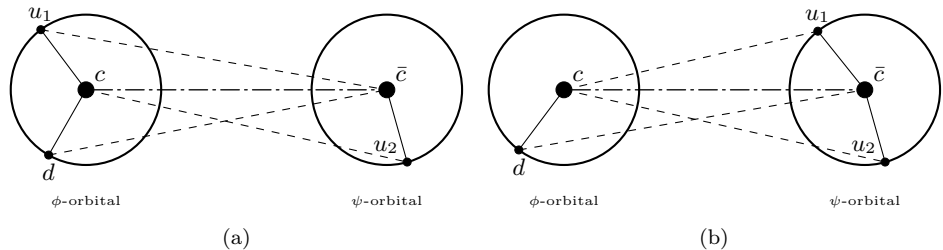
\begin{figure}[h]
\centering
    \subfigure[]{    
       \begin{tikzpicture}
\coordinate (A) at (0,0);
\coordinate (B) at (4,0);
\coordinate (C) at (-0.6,0.8);
\coordinate (D) at (-0.496,-0.868);
\coordinate (E) at (4.27,-0.961);
  \draw[thick] (A) circle (1cm);
  \node[below] at (0,-1.3) {\tiny $\phi$-orbital};
  \draw[thick] (B) circle (1cm);
    \node[below] at (4,-1.3) {\tiny $\psi$-orbital};
  \draw[dotted, line cap=round, line width=0.5pt, dash pattern=on 1pt off 2pt on 7pt off 2pt] (A)--(B);
   \draw (A)--(D);
    \draw (A)--(C);
     \draw (B)--(E);
      \draw[dashed] (C)--(B);
      \draw[dashed] (D)--(B);
      \draw[dashed] (A)--(E); 
  \filldraw[black] (A) circle (3pt) node[anchor=south west] {$c$};
   \filldraw[black] (B) circle (3pt) node[anchor=south west] {$\bar c$};
     \filldraw[black] (C) circle (1.5pt) node[above] {$u_1$};
     \filldraw[black] (D) circle (1.5pt) node[below] {$d$};    
        \filldraw[black] (E) circle (1.5pt) node[anchor=south east] {$u_2$};    
\end{tikzpicture}
    }
    \subfigure[]{
     \begin{tikzpicture}
\coordinate (A) at (0,0);
\coordinate (B) at (4,0);
\coordinate (C) at (-0.6,-0.8);
\coordinate (D) at (3.375,0.78);
\coordinate (E) at (4.27,-0.961);
  \draw[thick] (A) circle (1cm);
  \node[below] at (0,-1.3) {\tiny $\phi$-orbital};
  \draw[thick] (B) circle (1cm);
    \node[below] at (4,-1.3) {\tiny $\psi$-orbital};
  \draw[dotted, line cap=round, line width=0.5pt, dash pattern=on 1pt off 2pt on 7pt off 2pt]  (A)--(B);
   \draw[dashed] (A)--(D);
    \draw (A)--(C);
     \draw (B)--(E);
      \draw[dashed] (C)--(B);
      \draw (D)--(B);
      \draw[dashed] (A)--(E); 
  \filldraw[black] (A) circle (3pt) node[anchor=south west] {$c$};
   \filldraw[black] (B) circle (3pt) node[anchor=south west] {$\bar c$};
     \filldraw[black] (C) circle (1.5pt) node[below] {$d$};
     \filldraw[black] (D) circle (1.5pt) node[above] {$u_1$};    
        \filldraw[black] (E) circle (1.5pt) node[anchor=south east] {$u_2$};    
\end{tikzpicture}
    }
    \caption{\footnotesize{Two different possibilities for quark occupation distributed in two orbitals. In (a) two quarks sit in the $c$ orbital (denoted by $\phi$); in (b) two quarks sit in the $\bar c$ orbital (denoted by $\psi$). 
    Full lines indicate interactions that make the orbitals, dotted lines indicate additional interactions taken into account to first order perturbation theory. The dot-dashed line represents the Born-Oppenheimer potential.  For consistency, as discussed in  Sect.~\ref{consist}, light quarks in the same orbital must have the same interaction with $c$ and $\bar c$.} }
 \label{2orbit} 
  \end{figure}
 
 \item One can also consider a {\it two orbitals} scheme with two separate Schr\"odinger equations:  orbital $q-c$ (denoted by $\phi$) and orbital $\bar c$ (denoted by $\psi$). In Fig.~\ref{2orbit} (a) two light quarks sit in $\phi$ and one in $\psi$, the opposite in Fig.~\ref{2orbit} (b). 
 We have two distinct possibilities for the ground state
\bea
&&\Psi_0^{(a)}=\phi_0(x_d)\phi_0(x_{u_1})\psi_0(x_{u_2})
\notag\\
&&\Psi_0^{(b)}=\phi_0(x_d)\psi_0(x_{u_1})\psi_0(x_{u_2})
\eea
\end{enumerate}
{\bf \emph{A consistency condition.}} QED charges of protons and electrons are fixed constants. This is not the case for  two-body QCD charges, which depend on the superpositions of the relative color representations in which the pair occurs, as indicated in Sect.~\ref{due}. This leads to a consistency condition, namely that:
{\it quarks of different flavor in the same orbital (as in Fig.~1) must share the same QCD coupling to the heavy quarks at the center of the orbital}
\be
\lambda_{{cq}}=\lambda_{{cq^\prime}}\qquad \lambda_{\bar c q}=\lambda_{\bar c q^\prime}\label{eqcoup} 
\ee
with possibly $\lambda_{cq}\neq \lambda_{\bar c q}$.

We shall see that this condition is not trivially satisfied in the pentaquark.

\section{Three light quark operators and Fermi statistics}\label{fstat}

{\bf \emph{Color singlet baryons.}}
 $S$-wave, color singlet baryons are fully antisymmetric under color exchange and fully symmetric under coordinate exchange. Therefore quark (flavour\,$\otimes$\,spin) must be symmetric.  
 
 If we summarise spin and flavour quantum numbers with representations of SU(6)\,$\supset$\,SU(3)$_f\, \otimes$\, SU(2)$_{\rm spin}$, this is the ${\bm{56}}$ representation of non-relativistic SU(6)~\cite{Gursey:1992dc}, with content
\be
{\bm{56}:}~({\bm{8}},1/2) \oplus  ({\bm{10}}, 3/2)
\ee

A different case is that of excited, negative parity baryons: color is fully antisymmetric, but coordinates are in a mixed state (two quarks in $S$-wave and one quark in $P$-wave), see Ref.~\cite{isgur,roberts}.  The three quarks must form a {\it mixed-symmetry} representation in  flavour and spin, to obtain full symmetry when combined with coordinates. In SU(6) language, negative parity baryons are in the ${\bf {70}}$ representation, which decomposes as
\be
{\bm{70}:}~({\bm 1},1/2)\oplus ({\bm 8},3/2) \oplus ({\bm 8},1/2) \oplus ({\bm {10}},1/2). \label{seventy}
\ee
The third three-quark SU(6) representation, the fully antisymmetric 
\be
{\bf{20}}=({\bf 1},3/2)\oplus ({\bf 8},1/2)
\ee
is forbidden by Fermi statistics, for both ground state and $P$ wave baryons.

{\bf \emph{Pentaquarks.}}
$S$-wave pentaquarks have the three light quarks in color octet, distributed in one or more orbitals. Total antisymmetry under quark exchange required by Fermi statistics may be reached in different ways, summarized in Tab.~\ref{class}.

\begin{table}[htb!]
\centering
   \begin{tabular}{|c|c|c|c|}
     \hline
{\footnotesize Colour}&  {\footnotesize Coordinates} & {\footnotesize$SU(6)$}& {\footnotesize Notes} \\
 \hline
 {\footnotesize $M$}& {\footnotesize $S$} &\footnotesize{$\bm{70}$ $(M)$} & {\footnotesize Colour $\otimes SU(6)$: $A$}\\
 \hline
 {\footnotesize $M$}& {\footnotesize $M$}  &\footnotesize{$\bm{56}$ $(S)$}& {\footnotesize Colour $\otimes$ Coordinates: $A$}\\
 \hline
{\footnotesize $M$} & {\footnotesize $M$}&\footnotesize{$\bm{20}$} $(A)$ & {\footnotesize Colour $\otimes$ Coordinates: $S$}\\
 \hline
  {\footnotesize $M$} & {\footnotesize $M$}& \footnotesize{$\bm{70}$ $(M)$}
   & {\footnotesize Colour $\otimes$ Coordinates: $M$}\\
  \hline
\end{tabular}
 \caption{\footnotesize {To get a totally anti-symmetric  $(A)$ state, we need $M\times M\to A$ or $M\times M\to S$ depending on the SU(6) or coordinate entries. In the last row we require that the mixed symmetries for color and coordinate are composed in a mixed symmetry which, combined with spin-flavor SU(6), makes a totally anti-symmetric state.}}
\label{class}
\end{table}

The simplest possibility~is to assume complete symmetry of the coordinates. In this case, full antisymmetry under quark exchange requires the light quark complex to have  mixed symmetry under spin and flavour exchange, to be combined with colour to a totally antisymmetric state, first row of Tab.~\ref{class}: the three light quarks must form a ${\bm{70}}$ representation~\cite{Santopinto:2016pkp}, with the flavour-spin content reported in Eq.~\eqref{seventy}.

However, one needs to take into account the consistency condition stated in Sect.~\ref{consist}. 

\begin{enumerate}
\item In Born-Oppenheimer parlance, symmetry under coordinate exchanges means that light quarks populate a single orbital, Fig.~\ref{ellissi}. Accordingly, they must share the same QCD coupling to $c$ and to $\bar c$  
\be
\lambda_{cu}=\lambda_{cd}\qquad \lambda_{\bar c u}=\lambda_{\bar c d}
\ee
with possibly $\lambda_{cq}\neq \lambda_{\bar c q}$. 
An explicit calculation for the case of the $({\bm {8}}, 3/2)\subset {\bm{70}}$, see App.~\ref{oneorb}, shows however that the couplings of $d$ and $u$ quarks are different: {\it light quarks in a color octet cannot populate a single orbital}.
\item The next simplest possibility, in analogy with $L=1$ baryons,  is to arrange light quarks in two orbitals~\footnote{The situation is analogous to the molecular structure of the ion $H_2^-=2 p ~3 e^- $}, one around $c$ and the other around $\bar c$, Fig.~\ref{2orbit}, (a) or (b). 
Combining color mixed-symmetry with coordinate mixed-symmetry, one may obtain color-coordinate symmetry of type $A$, $S$, or $M$. To obey Fermi statistics, these possibilities must be associated to flavor-spin symmetries  of type $S$, $A$ or $M$, respectively, namely to the SU(6) representations ${\bm {56}}$, ${\bm {20}}$ or ${\bf {70}}$ (second to fourth rows of Tab.~\ref{class}).
\end{enumerate}

 We analyse the three cases in  Sect.~\ref{quattro}, and show that only ${\bm {56}}$ and ${\bm {20}}$ obey the consistency condition stated in Sect.~\ref{consist}.

\section{Octets and decuplets with light quarks in two orbitals} \label{quattro}

We give in this Section the explicit form of the light quarks operators corresponding to the $SU(6)$ representations {\bf{56}}, {\bf{20}} and {\bf{70}},  with quarks distributed in two orbitals denoted by  $\phi$ and $\psi$ and mixed-symmetry in the coordinates (see~\eqref{coomixed} and rows 2 to 4 in Tab.~\ref{class}). 

We give a complete discussion of the spin 1/2 octet, $({\bf 8},1/2)$ of ${\bm{56}}$ and report the results for the other cases, referring to App.~\ref{quattroA} for details.

In the following, we use the following abbreviated notations
\bl
&[q_1,q_2]q_3  \Leftrightarrow ((q_1,q_2)_{\bar{\bm 3}_c} q_3)_{\bm 8_c}\notag\\
&(q_1,q_2)q_3  \Leftrightarrow ((q_1,q_2)_{{\bm 6}_c} q_3)_{\bm 8_c}\notag\\
&[q_1,q_2]_{s}\vee (q_1,q_2)_{s} \Leftrightarrow (q_1,q_2)_{\bar{\bm 3}_c,S=s}\vee  (q_1,q_2)_{\bm 6_c,S=s}\notag\\
&\{[q_1,q_2] q_3\}_s \vee \{(q_1,q_2) q_3\}_s  \Leftrightarrow ([q_1,q_2] q_3)_{S=s}\vee  ((q_1,q_2) q_3)_{S=s}
\el
Whenever we write $(c\bar c)$ we mean color octet $(c\bar c)_{\bm 8}$

\subsection{The spin 1/2 octet of {\bf{56}}} 
Following Eqs.~\eqref{composizione}, the fully antisymmetric wave function in color and coordinate is
\be
{\cal P}=(\bar c c)\times \frac{1}{\sqrt{2}} [M^\lambda N^\rho - M^\rho N^\lambda]
=(c\bar c)\times \frac{1}{\sqrt{2}} \left\{ (q_1,q_2) q_3 \, N^\rho-[q_1,q_2]q_3 \,N^\lambda\right\}\label{fullyAS}
\ee
For color ${\bf 6}$ or $\bar{\bm 3}$ , when $q_1$ and $q_3$ are in $\phi$ and $q_2$ in $\psi$ we introduce the abbreviation
\bea
&&(q_1,q_2)q_3\, \phi(1)\psi(2)\phi(3)\equiv (q^\phi_1,q^\psi_2) q_3^\phi\notag\\
&&[q_1,q_2]q_3\, \phi(1)\psi(2)\phi(3)\equiv [q^\phi_1,q^\psi_2] q_3^\phi
\eea
We obtain
\be
{\cal P}=(\bar c c)\times \frac{-n}{\sqrt{2}}\left\{\frac{1}{\sqrt{2}}\left((q^\phi_1,q^\psi_2) q_3^\phi-(q^\psi_1,q^\phi_2) q_3^\phi\right) + \frac{1}{\sqrt{6}} \left(
[q^\phi_1,q^\psi_2] q^\phi_3+[q^\psi_1,q^\phi_2] q^\phi_3-2[q^\phi_1,q^\phi_2] q^\psi_3\right)\right\}\label{penta56}
\ee
Antisymmetry of Eq.~\eqref{penta56} under the exchange $(1,2)$ is explicit. To check the case $(1,3)$ one has to use Tab.~\ref{uno2}. A tedious but direct calculation shows that indeed
\bea
&&(1,3){\cal P}=-{\cal P}\label{compas}
\eea
Using Tab.~\ref{uno2}, we bring all $q^\phi$ inside the parentheses and obtain the more symmetric form (dropping an irrelevant overall sign)
\be
{\cal P}=(\bar c c)\times \frac{n}{\sqrt{3}}~\Big\{[q_1^\phi,q_2^\phi] q_3^\psi + \text{cyclic permutations in 1,2,3}\Big\}\label{cyclic56}
\ee
We have to combine Eq.~\eqref{cyclic56} with the flavour-spin fully symmetric expression obtained from the mixed-symmetry templates, Eqs.~\eqref{flavoct} and~\eqref{spin1/2} in App.~\ref{templates} to obtain
\be
S({\bm {56}})=\frac{1}{\sqrt{2}}[F^\rho \chi^\rho+F^\lambda \chi^\lambda] \label{simm56}
\ee 
{\bf \emph{The $\bm {d_3=d}$ condition.}} We may use \eqref{cyclic56}, \eqref{simm56} and Tab.~\ref{uno2} to bring the $d_i$ in the same position.One obtains the superposition of three replicas which are cyclical permutations of the assignments of $u$ and $d$ flavours to quarks $1,~2,~3$. In practical calculations, we may decide that the $d$ flavour is in quark $q_3$, keep the term containing $d_3=d$ and discard the rest. To obtain a normalised state we have to multiply  by $\sqrt{3}$. 

In a way, the condition 
\be
d_1=d_2=0\qquad d_3=d \label{gaugefix}
\ee
is analogous to fixing the gauge in a gauge invariant theory and restrict to one of the (infinitely many) orbits generated by gauge transformations. In our case, the group is the group of quark permutations and there are only three orbits.

Applying the condition $d=d_3$ to \eqref{simm56}  only the second term in \eqref{simm56} is non vanishing and we obtain~\footnote{Reassuringly, one verifies that the choice: $d_1=d$ gives the equivalent result: $S({\bm {56}})=-1/\sqrt{3}\{\lceil u_2, u_3\rfloor d\}_{1/2}$.}
\be
S({\bm {56}})=-\frac{1}{\sqrt{3}}~\{\{u_1, u_2\}_1 d\}_{1/2}\label{esse56}
\ee
or more explicitely
\be
S({\bm {56}})=\frac{1}{\sqrt{3}}\Big[\frac{u_1^\uparrow u_2^\downarrow+u_1^\downarrow u_2^\uparrow}{\sqrt{2}}~d^\uparrow-\sqrt{2}~u_1^\uparrow u_2^\uparrow d^\downarrow\Big]  \label{def1/2}
\ee

Combining \eqref{esse56} and \eqref{penta56}, the operator that creates ${\cal P}_{\bm {56}}$ octet takes the proton-like form (the extension to the other members of the octet is given in App.~\ref{fulloct56})
\bea
{\cal P}^{\bm { 56}}_{\bf 8,1/2}&=&\sqrt{3}~ {\cal P}\otimes S({\bm {56}})= \notag\\
&=&(\bar c c)\times \frac{n}{\sqrt{2}}\left\{\frac{1}{\sqrt{2}}\left((u^\phi_1,u^\psi_2 )_1 d^\phi -(u^\psi_1,u^\phi_2)_1 d^\phi \right) + \frac{1}{\sqrt{6}} \left(
[u^\phi_1,u^\psi_2]_1 d^\phi+[u^\psi_1,u^\phi_2]_1d^\phi-2[u^\phi_1,u^\phi_2]_1d^\psi\right)\right\}\label{primo56}
\eea

It is convenient to have an expression which puts together quarks that belong to the same orbital. This is obtained by using \eqref{esse56} with \eqref{cyclic56} and further expressing the wave function in terms of the total spin of the pair inside orbital $\phi$
\bea
{\cal P}^{\bm { 56}}_{\bf 8,1/2}&=&\sqrt{3}~ {\cal P}\otimes S({\bm {56}})=\notag\\
&=&(\bar c c)\times \frac{n}{\sqrt{3}}\Big\{[u_1^\phi,u_2^\phi]_1 d^\psi -\frac{1}{2}[d^\phi,u_1^\phi]_1 u_2^\psi -\frac{1}{2}[ u_2^\phi,d^\phi]_1u_1^\psi
-\frac{\sqrt{3}}{2}[d^\phi,u_1^\phi ]_0 u_2^\psi+\frac{\sqrt{3}}{2}[u_2^\phi,d^\phi ]_0 u_1^\psi~\Big\}
\label{orbitalsep56}
\eea

Color couplings of $d^\phi$ and $d^\psi$ can be easily calculated from the above formulas
\bea
&&\lambda_{d^\phi c}=\lambda_{u_1^\phi c}=-\frac{7}{18}\qquad \lambda_{d^\phi \bar c}=\lambda_{u_1^\phi \bar c}= -\frac{2}{18}\notag \\
&&\lambda_{d^\psi c}=\lambda_{u_1^\psi c}= -\frac{2}{18}\qquad  \lambda_{d^\psi \bar c}=\lambda_{u_1^\psi \bar c}=-\frac{7}{18}\label{56coup}
\eea 
This operator describes a state with the orbitals, $\phi$ and $\psi$, populated according to
\be
 \phi( qq), \psi (q) \notag 
\ee
We have a second possibility where $\psi$ has occupation number two. In total 
\be
\begin{aligned}
\label{summuno56} 
& {\bm A}({\bm {56}})  &   & c-\phi( qq)&& \bar c- \psi(q) & & {\rm Fig.~}\ref{2orbit} (a)\notag\\
& {\bm B}({\bm {56}})    &  &  c- \phi(q) &&  \bar c- \psi( qq)  & &  {\rm Fig.~}\ref{2orbit} (b)\notag
\end{aligned}
\ee
The first possibility has lower energy, $-7/18<-2/18$,  in the one gluon exchange approximation.
\vskip01cm

\subsection{The spin 3/2 decuplet of ${\bf 56}$.}

We register the case with $I_3=+1/2$
\bea
&&{\Delta}^{+}_{10}=(\bar c c)\times \frac{n}{\sqrt{3}}  \Big\{\{[u^\phi_1, u^\phi_2]d^\psi\}_{3/2}+\{[d^\phi, u^\phi_1]u_2^\psi\}_{3/2}+\{[u_2^\phi, d^\phi]u_1^\psi\}_{3/2} \Big\} \label{deltap56}
\eea
Color couplings $u_3^{\phi,\psi}-c/\bar c$ are the same as in \eqref{56coup}.

\subsection{The spin 1/2 octet of {\bf{20}}}

We find
\bea
&&{\cal P}^{\bm{20}}_{\bm{8,1/2}}=(\bar c c)\times \frac{n}{\sqrt{2}} \Big\{ \frac{ [ u_1^\phi, u_2^\psi]_0 d^\phi- [ u_1^\psi, u_2^\phi]_0 d^\phi }{\sqrt{2}}-\frac{( u_1^\phi,u^\psi_2)_0 d^\phi+( u_1^\psi,u^\phi_2)_0 d^\phi-2(u_1^\phi,u^\phi_2)_0 d^\psi}{\sqrt{6}}\Big\}\label{primo20}
\eea
or, equivalently
\bea
{\cal P}^{\bm{20}}_{\bm{8,1/2}}&=&(\bar c c)\times  \frac{n}{\sqrt{3}} \Big[ \{( u_1^\phi,u_2^\phi)_0 d^\psi\}_{1/2}-\frac{1}{2}\{(d^\phi, u_1^\phi)_0 u_2^\psi \}_{1/2}-\frac{1}{2}\{(  d^\phi, u_2^\phi)_0 u_1^\psi \}_{1/2}+\notag \\
&-&\frac{\sqrt{3}}{2}\{( d^\phi,u_1^\phi)_1u_2^\psi \}_{1/2}+\frac{\sqrt{3}}{2}\{ (d^\phi, u_2^\phi)_1u_1^\psi \}_{1/2}\Big]
\label{orbitsep20}
\eea
 
Not surprisingly, \eqref{orbitsep20} is obtained from \eqref{orbitalsep56} with {\it two} simultaneous exchanges of the symmetry characters
\be
\bar{\bm 3}_c \to \bm 6_c\qquad \text{diquark~spin:}~1\rightleftarrows 0\qquad \text{isospin~unchanged}
\ee
Thus, both octets have {\it the same antisymmetry} under total exchange of quark quantum numbers, as required by Fermi statistics. 

We find the couplings
\bea
&&\lambda_{ d^\phi c}=\lambda_{ u_1^\phi c}=-\frac{5}{18}\qquad  \lambda_{d^\phi {\bar c}}= \lambda_{u_1^\phi {\bar c}}=-\frac{10}{18}\notag \\
&&\lambda_{d^\psi c}=\lambda_{u_1^\psi c}=-\frac{4}{18}\qquad \lambda_{d^\psi {\bar c}}=\lambda_{u_1^\psi {\bar c}}=+\frac{1}{18}\notag
\eea

There is only one possible combination of orbitals, corresponding to Fig.~\ref{2orbit} (b)
\be
\begin{aligned}
\label{summ20}
& {\bm B}({\bm{20}})  &   & c- \phi(q) &&  \bar c- \psi( qq)  & & {\rm Fig.~}\ref{2orbit} (b)
\end{aligned}
\ee

\subsection{The spin 3/2 octet of {\bf{70}}}

We obtain
\bea
&&{\cal P}^{\bm {70}}_{\bm{8,3/2}}=n\frac{1}{\sqrt{2}}\Big\{\frac{(u^\phi_1,u^\psi_2) d^\phi-(u^\psi_1,u^\phi_2) d^\phi}{\sqrt{2}} 
-\frac{[u^\phi_1,u^\psi_2] d^\phi+[u^\psi_1,u^\phi_2] d^\phi -2[u^\phi_1,u^\phi_2] d^\psi }{\sqrt{6}}\Big\}\notag\\
\eea
$J_3=3/2$ all spins up. To bring quarks in the same orbital together, we use Tab.~\ref{uno2} to find
\bea
&&{\cal P}^{\bm {70}}_{\bm{8,3/2}}=n\frac{1}{\sqrt{2}}\times \Big\{ \frac{[u_1^\phi,d^\phi] u_2^\psi
+[d^\phi, u_2^\phi] u_1^\psi-2[u_1^\phi,u_2^\phi] d^\psi}{\sqrt{6}}-\frac{(u_1^\phi,d^\phi) u^\psi_2-(d^\phi, u_2^\phi) u_1^\psi}{\sqrt{2}}\Big\}
\eea

$d$ couplings are the same as $\bm{8,1/2}$ of $\bm{56}$
\bea
&&\lambda_{d^\phi c}=-\frac{14}{36}\qquad \lambda_{d^\phi \bar c}= -\frac{2}{18}\notag \\
&&\lambda_{d^\psi c}= -\frac{4}{36}\qquad \lambda_{d^\psi \bar c}=-\frac{7}{18}\notag
\eea 
but {\it $u$ couplings do not agree}. We find
\bea
&&\lambda_{u_1^\phi c} =-\frac{11}{36}\qquad \lambda_{u_1^\phi\bar c}=-\frac{8}{18}\notag\\
&& \lambda_{u_1^\psi c}=-\frac{7}{36}\qquad \lambda_{u_1^\psi\bar c}=-\frac{1}{18}\notag
\eea
 The conclusion is that the ${\bf {70}}$ is incompatible even with 2 orbitals. A similar disagreement is found between the couplings of the spin 1/2 octet of {\bf{70}}.

\section{Results and perspectives}  \label{result}

We summarize here our results, valid for the spin 1/2 octet of either ${\bm {56}}$ or ${\bm {20}}$.
\begin{enumerate}
\item Combining the spin 1/2 of light quarks with $c\bar c$ total spin $S_c=s_c+s_{\bar c}=0,1$, we obtain {\it three pentaquark octets} with spin compositions: 2$({\bm 8},~1/2)+({\bm 8},~3/2)$.
\item All $S=0$ states can decay into a final state containing a $J/\psi+p$. Conservation of $S_c$, which would forbid the final $J/\psi$ for $S_c=0$, is broken by light-heavy hyperfine interactions in the two orbitals. The same applies to strange pentaquarks.
\item We expect therefore {\it {three pentaquark lines}} for both $S=0$ and $S=-1$, with
\be
{\cal P}_{(S=0)}\to J/\psi +p \qquad {\cal P}_{(\Lambda,S=-1)} \to J/\psi +\Lambda
\ee
\item Other predicted pentaquarks are
\bea
&&{\cal P}^+_{(\Sigma,S=-1)} \to J/\psi +\Sigma^+ \to J/\psi +p +\pi^0\notag \\
&&{\cal P}_{(\Xi,S=-2)}^- \to J/\psi + \Xi^- \to J/\psi +\Lambda+ \pi^-
\eea
\item The two alternatives,  ${\bm {56}}$ or ${\bm {20}}$, are distinguished by presence or absence of pentaquarks decaying into spin 3/2 resonances, e.g.: ${\cal P}_{(\Delta,S=0)}^{++} \to J/\psi + \Delta^{++}\to J/\psi + p+\pi^+$, which applies to ${\bm {56}}$ only.
\item Taking one-gluon exchange couplings, Sect.~\ref{quattro}, one would conclude for a ground state in the {\bf {20}} representation. However, the one gluon exchange approximation is not that precise and we consider both cases equally plausible for the ground state, to be decided by presence or absence of  the SU(3)$_f$ decuplet.
\end{enumerate}

Before closing, we comment on the possibility to compute the mass spectrum of pentaquarks.
In our two-orbitals scheme there are three ground state candidates, namely
\be
\begin{aligned}
& \bm{56}^{(a)}  &   &\text{ Fig.~\ref{2orbit} (a)}  && (c q q^\prime)^\phi(\bar c q^{\prime\prime})^\psi \\
& \bm{56}^{(b)}  &   &\text{ Fig.~\ref{2orbit} (b)}  && (c q )^\phi(\bar c  q^\prime q^{\prime\prime})^\psi \\
& \bm{20}^{(b)}  &   &\text{ Fig.~\ref{2orbit} (b)}  && (c q )^\phi(\bar c  q^\prime q^{\prime\prime})^\psi \notag
\end{aligned}
\ee

We start from the first case, $\bm{56}^{a}$.
 
A glance at Eq.~\eqref{tf2} shows that orbitals at large $R$ are in triality zero, color configurations ${\bf {8}}_c-{\bf { 8}}_c$ or ${\bf {1}}_c-{\bf {1}}_c$. 
As argued in~\cite{Bali:2000gf}, soft gluons may screen colors in the ${\bf {8}}_c-{\bf { 8}}_c$ configuration and the BO potential vanishes at infinity~\cite{noiBO}. For tetraquarks this behaviour is  supported by the recent lattice QCD calculation of~\cite{Bicudo:2012qt}. The pentaquark goes asymptotically into a superposition of charmed baryon-anticharmed meson states. The pentaquark at intermediate distances, is a kind of molecular state bound by QCD interactions. In this case, the mass spectrum would be fully computable, similarly to the doubly charm tetraquark mass~\cite{Karliner:2013dqa,Eichten:2017ffp,Maiani:2022qze}.

Orbitals in the second case, $\bm{56}^{b},~\bm{20}^b$, are shown in Fig.~\ref{2orbit}(b). The configuration reminds closely the compact pentaquark: $[cu](\bar c [ud])$ proposed in~\cite{noiBO}. Orbitals are in color confined configurations~${\bf {\bar 3}}_c-{\bf { 3}}_c$ or ${\bf {6}}_c-{\bf {\bar 6}}_c$, see \eqref{tf1}. We must add to the BO potential a string potential rising to infinity for $R\to \infty$ ~\cite{noiBO}. As it happens for charmonia, pentaquark masses contain one undetermined constant and one can compute only mass differences with respect to ground state, i.e. those due to the hyperfine interactions.


\vspace{2truecm}
\appendix

\section{The case of one orbital} \label{oneorb}
For the spin 3/2 octet ($J_3=1/2,~J=3/2$) we factorise the symmetric spin part (i.e. assume all spins up and ignore spin altoghether) and use \eqref{flavoct} 
\bea 
 {\cal P}_{\bm{8,3/2,\text{\bf 1-orbit}}}^{\bm{70}}&=&\frac{1}{\sqrt{2}}[M^\rho F^\lambda-M^\lambda F^\rho]\otimes \chi^S=\notag\\
 &=&-\frac{1}{2}\Big\{\frac{1}{\sqrt{3}}\left( [ u_1,d_2]u_3+[d_1,u_2] u_3 -2 [ u_1, u_2] d_3\right)+\left((u_1,d_2) u_3-(d_1,u_2) u_3\right)\Big\}\label{b8threehalffull}
 \eea

{\bf\emph{Fixing the gauge $\bm {d=d_3}.$}} As illustrated in Sect.~\ref{quattro}, we may use the gauge condition that $d$ flavour is in quark $q_3$, keep the term containing $d_3=d$ and discarding the rest. To obtain a normalised state we multiply  by $\sqrt{3}$. 


Applied to \eqref{b8threehalffull} the gauge condition $d_3=d$ leads to the simple result (all spins up):
\be
{\cal P}_{\bm{8,3/2,\text{\bf 1-orbit}}}^{\bm{70}}(d_3=d)=[u_1, u_2] d\label{b8threehalfred}
\ee

We find
\bea
&& \lambda_{dc}=\frac{2}{3}\left(- \frac{2}{3}\right) +\frac{1}{3}\left(\frac{1}{3}\right)=-\frac{2}{6}\notag\\
&& \lambda_{d\bar c}=\frac{8}{9}\left(-\frac{4}{3}\right)+\frac{1}{9}\left(+\frac{1}{6}\right)=-\frac{7}{6}
\eea

Next, using Tab.~\ref{uno2}, we find
\bea
&&[ u_1, u_2] d =+\frac{1}{2}[d, u_2]u_1-\frac{\sqrt{3}}{2} (d, u_2) u_1
\eea
and 
\bea
&& \lambda_{u_1c}=\frac{1}{4}\left[\frac{2}{3}\left(-\frac{2}{3}\right)+\frac{1}{3}\left(+\frac{1}{3}\right)\right]+\frac{3}{4}\left(-\frac{2}{3}\right)=-\frac{7}{12}\notag\\
&& \lambda_{u_1\bar c}=\frac{1}{4}\left[\frac{8}{9}\left(-\frac{4}{3}\right)+\frac{1}{9}\left(+\frac{1}{6}\right)\right]+\frac{3}{4}\left(+\frac{1}{6}\right)=-\frac{2}{12}
 \eea

 \section{Templates of mixed and symmetric representations of $S_3$} \label{templates}

 For color we use the definitions
\bea
&& M^\lambda=|(q_1,q_2)_{\bm 6}\, q_3\rangle_{\bm 8}\label{simmA}\\
&& M^\rho=|(q_1,q_2)_{\bar {\bm 3}}\, q_3 \rangle_{\bm 8}\label{asimmA}
\eea
In the case of flavor, octet or decuplet, we have 

\bea
&&F^\rho=\frac{1}{\sqrt{2}} (u_1d_2-d_1u_2)u_3\notag\\
&& F^\lambda=-\frac{1}{\sqrt{6}}\left[ (u_1d_2+d_1u_2)u_3 -2 u_1 u_2 d_3\right] ~{\rm (octet)}\label{flavoct}\\
&&F^S=\frac{1}{\sqrt{3}}(d_1u_2u_3+u_1d_2u_3+u_1u_2d_3)~{\rm (decuplet)} \label{flavdec}
\eea
and similarly for spin
\bea
&& \chi^\rho=\frac{1}{\sqrt{2}} (\uparrow \downarrow-\downarrow \uparrow )\uparrow\notag\\
&& \chi^\lambda=-\frac{1}{\sqrt{6}}\left[(\uparrow \downarrow+ \downarrow \uparrow )\uparrow-2(\uparrow \uparrow) \downarrow\right] ~(\rm spin~1/2) \label{spin1/2}\\
&& \chi^S=\frac{1}{\sqrt{3}}(\uparrow \uparrow \downarrow+\uparrow \downarrow\uparrow + \downarrow\uparrow\uparrow)~(\rm spin~3/2)
\label{spin3/2}
\eea
For coordinates, $\phi,\psi$ refer to Born-Oppenheimer orbitals, Fig.~\ref{2orbit}, and
\bea
&&N^\rho=n~\frac{1}{\sqrt{2}}~\left[\phi(1)\psi(2)-\phi(2)\psi(1)\right]\phi(3)\notag\\
&&N^\lambda=-n~\frac{1}{\sqrt{6}}~\left\{[\phi(1)\psi(2)+\phi(2)\psi(1)]\phi(3)-2\phi(1)\phi(2)\psi(3)\right\}\label{coomixed}\\
&&n=\frac{1}{\sqrt{1-S^2}}\qquad\text{and}\qquad S=\int~d1~ \phi(1)\psi(1)<1, ~(\phi,~\psi =~{\rm real})\notag
\eea

 \section{Light quarks in two orbitals} \label{quattroA}
 
\subsection{The decuplet of ${\bf 56}$.}
The simplest case is to couple the state \eqref{penta56} with the $({\bf 10},3/2)$ of ${\bf 56}$, with wave function
\be
B_{10}= u^{\uparrow} u^{\uparrow}u^{\uparrow}
\ee
We find:
\bea 
{\Delta}^{++}_{10}&=&(\bar c c)\times \frac{n}{\sqrt{2}}\left\{ \frac{1}{\sqrt{2}}
(u^{\phi}_1u^{ \psi}_2-u^{ \psi}_1 u^{ \phi}_2) u^{ \phi}_3+
 \frac{1}{\sqrt{6}} \left(
[u^{ \phi}_1 u^{ \psi}_2+u^{ \psi}_1 u^{ \phi}_2] u^{ \phi}_3-2[u^{ \phi}_1 u^{ \phi}_2] u^{ \psi}_3\right)\right\}=\notag\\
&=&(\bar c c)\times \frac{n}{\sqrt{3}} \times \Big\{[u^\phi_1, u^\phi_2] u^\psi_3+[u_3^\phi, u^\phi_1] u_2^\psi+[u_2^\phi, u^\phi_3] u_1^\psi \Big\} \label{delta56}
\eea
Color couplings $u_3^{\phi,\psi}-c/\bar c$ are the same as in \eqref{56coup}. We register also the case with $I_3=+1/2$:
\bea
&&{\Delta}^{+}_{10}=(\bar c c)\times \frac{n}{\sqrt{3}}  \Big\{\{[u^\phi_1, u^\phi_2]d^\psi\}_{3/2}+\{[d^\phi, u^\phi_1]u_2^\psi\}_{3/2}+\{[u_2^\phi, d^\phi]u_1^\psi\}_{3/2} \Big\} 
\eea

\subsection{The octet of {\bf{20}}}

Following Eqs.~\eqref{composizione}, the fully symmetric wave function in color and coordinate is 
\bea
{\cal P}&=&(\bar c c)\times \frac{1}{\sqrt{2}} [M^\lambda N^\lambda + M^\rho N^\rho]= \notag \\
&=&(c\bar c) \times \frac{1}{\sqrt{2}} \left\{ (q_1,q_2) q_3\, N^\lambda+[q_1,q_2] q_3\, N^\rho\right\}
\eea
Introducing coordinate orbitals  as before, we obtain (subscript $S$ for symmetric)
\bea
{\cal P}_{S}=(\bar c c)_{\bm 8}\times\frac{n}{\sqrt{2}}
\times\left\{-\frac{1}{\sqrt{6}}\left((q^\phi_1,q^\psi_2) q_3^\phi+( q^\psi_1,q^\phi_2) q_3^\phi-2( q^\phi_1,q^\phi_2) q_3^\psi \right) + \frac{1}{\sqrt{2}} \left(
[q^\phi_1,q^\psi_2] q^\phi_3-[ q^\psi_1,q^\phi_2] q^\phi_3\right)\right\}\label{penta20}
\eea
Using Tab~\ref{uno2} we bring together quarks with the same orbitals and find, not surprisingly
\be
{\cal P}_{S}=(\bar c c)\times\frac{n}{\sqrt{3}}\Big[(q_1^\phi,q_2^\phi)q_3^\psi+\text{cyclic permutations in 1,2,3}\Big]\label{cyclic20}
\ee

 {\bf \emph{The full proton-like, octet state of  $\bm{20}$}}. 
We have to combine \eqref{penta20} or \eqref{cyclic20} with the flavour-spin fully antisymmetric expression obtained by combining the mixed symmetry templates in Sect.~\ref{tre}. Chosing again the gauge $d=d_3$ the antisymmetric combination is
\be
A({\bf{20}})=\frac{1}{\sqrt{3}}~\{ \{u_1, u_2\}_0 d \}_{1/2} \label{flavspin20}
\ee
with $\lfloor u_1, u_2\rceil$ meaning spin zero pair. Using \eqref{penta20}, we find
\be
{\cal P}^{\bm{20}}_{\bm{8,1/2}}= \sqrt{3} {\cal P}_{S} A({\bf{20}})
=(\bar c c)\times \frac{n}{\sqrt{2}} \Big\{ \frac{ [ u_1^\phi, u_2^\psi]_0 d^\phi- [ u_1^\psi, u_2^\phi]_0 d^\phi }{\sqrt{2}}-\frac{( u_1^\phi,u^\psi_2)_0 d^\phi+( u_1^\psi,u^\phi_2)_0 d^\phi-2(u_1^\phi,u^\phi_2)_0 d^\psi}{\sqrt{6}}\Big\}
\ee

We find the couplings repoprted before Eq.~\eqref{summ20}
\bea
&&\lambda_{ d^\phi c}=-\frac{5}{18}\qquad \lambda_{d^\phi {\bar c}}=-\frac{10}{18}\notag \\
&&\lambda_{d^\psi c}=-\frac{4}{18}\qquad \lambda_{d^\psi {\bar c}}=+\frac{1}{18}\notag
\eea
and one checks that the $u_1^{\phi,\psi}-c/\bar c$ couplings coincide with these.

There is only one possible combination of orbitals, corresponding to Fig.~\ref{2orbit} (b).
\be
\begin{aligned}
& {\bm B}({\bm{20}})  &   & c- \psi(q) &&  \bar c- \phi( qq) & & {\rm Fig.~}\ref{2orbit} (b)
\end{aligned}
\ee

{\bf \emph{Separating the orbitals.}} As for the ${\bm {56}}$, it is convenient to have an expression which puts together quarks that belong to the same orbital. This is obtained by using \eqref{cyclic20} with \eqref{flavspin20}  and rearranging quark spins so as to put into evidence the total spin inside each orbitalo. We obtain:
\bea
{\cal P}^{\bm { 20}}_{\bf 8,1/2}&=&\sqrt{3}~ {\cal P}\otimes A({\bm {20}})
=(\bar c c)\times  \frac{n}{\sqrt{3}} \Big\{\{( u_1^\phi,u_2^\phi)_0 d^\psi\}_{1/2}-\frac{1}{2}\{(d^\phi, u_1^\phi) u_2^\psi \}_{1/2}-\frac{1}{2}\{(  d^\phi, u_2^\phi)_0 u_1^\psi \}_{1/2}-\notag \\
&-&\frac{\sqrt{3}}{2}\{( d^\phi,u_1^\phi)u_2^\psi \}_{1/2}+\frac{\sqrt{3}}{2}\{ (d^\phi, u_2^\phi)u_1^\psi \}_{1/2}\Big\}
\label{orbitsep20x}
\eea

As observed above\eqref{orbitsep20x} is obtained from \eqref{orbitalsep56} with {\it two} simultaneous exchanges of the symmetry characters:
\be
\bar{\bm 3}_c \to \bm 6_c\qquad \text{diquark~spin:}~1\leftrightarrows 0
\ee

Going from \eqref{orbitalsep56} to \eqref{orbitsep20}, the diquark isospin remains unchanged.
Thus, both octets have {\it the same symmetry} under total exchange of quark quantum numbers, which is the required Fermi antisymmetry. 

\subsection{The spin 3/2 octet of {\bf{70}}}

{\bf \emph{Color and Coordinates mixed (M) wave functions.}}
Starting from Eqs.~\eqref{simm} and \eqref{asimm} together with \eqref{coomixed}, we have to construct
\bea
&&CC^\rho=\frac{1}{\sqrt{2}}~[M^\rho N^\lambda +M^\lambda N^\rho] \label{ccrho}\\
&&CC^\lambda=\frac{1}{\sqrt{2}}~[M^\rho N^\rho-M^\lambda N^\lambda] \label{cclam}
\eea
We find:
\bea
&&CC^\rho=n~\frac{1}{2\sqrt{3}}\Big\{\sqrt{3}\left((q^\phi_1,q^\psi_2) q^\phi_3-(q^\psi_1,q^\phi_2) q^\phi_3 \right)- 
[q^\phi_1,q^\psi_2] q^\phi_3-[q^\psi_1,q^\phi_2] q^\phi_3+2[q^\phi_1,q^\phi_2]q^\psi_3 \Big\}
\eea
and
\bea
&&CC^\lambda=n~\frac{1}{2\sqrt{3}}\Big\{\sqrt{3}\left([q^\phi_1,q^\psi_2] q^\phi_3-[q^\psi_1,q^\phi_2] q^\phi_3 \right)+ 
(q^\phi_1,q^\psi_2) q^\phi_3+(q^\psi_1,q^\phi_2) q^\phi_3-2(q^\phi_1,q^\phi_2)q^\psi_3 \Big\}
\eea
If we want the spin 3/2 octet we have to combine the above expressions with
\bea
F^\rho&=&\frac{1}{\sqrt{2}} (u_1d_2-d_1u_2)u_3\notag\\
F^\lambda&=&-\frac{1}{\sqrt{6}}\left[ (u_1d_2+d_1u_2)u_3 -2 u_1 u_2 d_3\right] ~{\rm (octet)}
\eea
 to obtain the full antisymmetric combination (all spins are up)
 \be
 A=\frac{1}{\sqrt{2}} [CC^\rho F^\lambda-CC^\lambda F^\rho]
 \ee

The gauge fixing condition $d_3=d$, Eq.~\eqref{gaugefix}, gives $F^\rho=0,~F^\lambda\propto u_1u_2d$ and one obtains

\bea
&&{\cal P}^{\bm {70}}_{\bm{8,3/2}}=n\frac{1}{\sqrt{2}}\Big\{\frac{(u^\phi_1,u^\psi_2) d^\phi-(u^\psi_1,u^\phi_2) d^\phi}{\sqrt{2}} 
-\frac{[u^\phi_1,u^\psi_2] d^\phi+[u^\psi_1,u^\phi_2] d^\phi -2[u^\phi_1,u^\phi_2] d^\psi }{\sqrt{6}}\Big\}\notag\\
\eea
$J_3=3/2$ all spins up. To bring quarks in the same orbital together, we use Tab.~\ref{uno2} to find:
\bea
&&{\cal P}^{\bm {70}}_{\bm{8,3/2}}=n\frac{1}{\sqrt{2}}\times \Big\{ \frac{[u_1^\phi,d^\phi] u_2^\psi
+[d^\phi, u_2^\phi] u_1^\psi-2[u_1^\phi,u_2^\phi] d^\psi}{\sqrt{6}}-\frac{(u_1^\phi,d^\phi) u^\psi_2-(d^\phi, u_2^\phi) u_1^\psi}{\sqrt{2}}\Big\}
\eea

$d$ couplings are the same as $\bm{8,1/2}$ of $\bm{56}$
\bea
&&\lambda_{d^\phi c}=-\frac{14}{36}\qquad \lambda_{d^\phi \bar c}= -\frac{2}{18}\notag \\
&&\lambda_{d^\psi c}= -\frac{4}{36}\qquad \lambda_{d^\psi \bar c}=-\frac{7}{18}\notag
\eea 
but {\it $u$ couplings do not agree}: we find
\bea
&&\lambda_{u^\phi_1c}=-\frac{11}{36}\qquad \lambda_{u^\phi_1\bar c}=-\frac{8}{18}\notag\\
&& \lambda_{u^\psi_1c}=-\frac{7}{36}\qquad \lambda_{u^\psi_1\bar c}=-\frac{1}{18}\notag
\eea
 The conclusion is that the ${\bf {70}}$ is incompatible even with 2 orbitals. A similar disagreement is found between the couplings of the spin 1/2 octet of {\bf{70}}.

\section{The full spin 1/2 octet of {\bf{56}}} \label{fulloct56}
We start from the proton wave function \eqref{orbitalsep56} (for the sake of clarity, we indicate explicitly the spin of diquarks in the squared parentheses)
\be
{P}=\frac{1}{\sqrt{3}} \Big\{[ u_1^\phi,u_2^\phi]_1 d^\psi-\frac{1}{2}[d^\phi,u_1^\phi]_1 u_2^\psi -\frac{1}{2} [ u_2^\phi,d^\phi]_1 u_1^\psi
-\frac{\sqrt{3}}{2}[ d^\phi,u_1^\phi ]_0 u_2^\psi+\frac{\sqrt{3}}{2}[ u_2^\phi,d^\phi ]_0 u_1^\psi~\Big\}
\label{proton56}
\ee
\begin{itemize}
\item Replacing $d\to s$ we obtain the $\Sigma^+$
\be
\Sigma^+=\frac{1}{\sqrt{3}} \Big\{[ u_1^\phi,u_2^\phi]_1 s^\psi-\frac{1}{2}[s^\phi,u_1^\phi]_1 u_2^\psi -\frac{1}{2} [ u_2^\phi,s^\phi]_1 u_1^\psi
-\frac{\sqrt{3}}{2}[ s^\phi,u_1^\phi ]_0 u_2^\psi+\frac{\sqrt{3}}{2}[ u_2^\phi,s^\phi ]_0 u_1^\psi~\Big\}
\ee
\item The substitution $u\to d$ leads to $\sqrt{2}\Sigma^0$ with ($I^-$ is the isospin lowering operator):
\bea
{\Sigma^0}&=&\frac{I^-\Sigma^+}{\sqrt{2}}=\frac{1}{\sqrt{3}}\times \Big\{\frac{[ d^\phi,u^\phi]_1 s^\psi+[ u^\phi,d^\phi]_1 s^\psi}{\sqrt{2}}-
\frac{1}{2}\Big[\frac{    [s^\phi,d^\phi]_1 u^\psi+[s^\phi,u^\phi]_1 d^\psi}{\sqrt{2}}+\frac{ [ s^\phi, d^\phi]_1 u^\psi +[ s^\phi, u^\phi]_1 d^\psi}{\sqrt{2}}\Big]+\notag \\
&-&\frac{\sqrt{3}}{2}\Big[\frac{    [ s^\phi,d^\phi ]_0 u^\psi-[d^\phi ,s^\phi ]_0 u^\psi  }{\sqrt{2}}+\frac{    [ s^\phi,u^\phi ]_0 d^\psi-[u^\phi ,s^\phi ]_0 d^\psi  }{\sqrt{2}}\Big]~\Big\}
\label{Sigz56}
\eea


 \item The rising operator of $V$spin in octet space (corresponding to $u\to s$)  is represented by the matrix
\be
V^+=\begin{pmatrix} 0&0&0\\0&0&0\\1&0&0 \end{pmatrix}
\ee
and it acts on the octet baryon matrix $B$ according to the commutator: $B\to \big[V^+,B\big]$. Specialising to the proton, we find
\be
 \frac{\big[V^+,P\big]}{\sqrt{2}}=\Sigma^0_V=-\frac{\sqrt{3}}{2}\Lambda-\frac{1}{2}\Sigma^0.    \label{vspin}
 \ee
 Application of $V^+=u\to s$ brings  the state~\eqref{proton56} into $\sqrt{2}\Sigma^0_V$:
\bea
\Sigma_V&=&\frac{1}{\sqrt{3}}\times \Big\{\frac{[ s^\phi,u^\phi]_{1}d^\psi+[u^\phi,s^\phi]_1d^\psi}{\sqrt{2}}-
\frac{1}{2}\Big[\frac{[d^\phi,s^\phi]_1 u^\psi+ [s^\phi,d^\phi]_1 u^\psi }{\sqrt{2}} +
\frac{[d^\phi,u^\phi]_1 s^\psi+[u^\phi,d^\phi]_1 s^\psi }{\sqrt{2}}\Big]-\notag \\
&-&\frac{\sqrt{3}}{2}\Big[\frac{[d^\phi,s^\phi]_0 u^\psi -[s^\phi,d^\phi]_0 u^\psi }{\sqrt{2}}-
\frac{[u^\phi,d^\phi]_0 s^\psi -[d^\phi,u^\phi]_0 s^\psi}{\sqrt{2}}
\Big]~\Big\}
\label{VP56}
\eea
Using~\eqref{vspin}, we find
\bea
\Lambda&=&-\frac{1}{\sqrt{3}}\big(2\Sigma^0_V+\Sigma^0)=
\frac{1}{\sqrt{3}}\times\Big\{\frac{-\sqrt{3}}{2}\Big[\frac{[s^\phi,u^\phi ]_1 d^\psi +[u^\phi,s^\phi]_1d^\psi }{\sqrt{2}}-\frac{[s^\phi,d^\phi]_1 u^\psi +[d^\phi,s^\phi ]_1 u^\psi }{\sqrt{2}}\Big]-\notag\\
&-&\frac{1}{2}\Big[\frac{[s^\phi,d^\phi]_0 u^\psi - [ d^\phi,s^\phi]_0 u^\psi }{\sqrt{2}}-
\frac{[s^\phi,u^\phi]_0 d^\psi -[u^\phi,s^\phi]_0 d^\psi }{\sqrt{2}}\Big]
-\frac{[ u^\phi,d^\phi]_0 s^\psi -[d^\phi,u^\phi]_0 s^\psi }{\sqrt{2}}\Big\} \label{Lamb56}
\eea
which, correctly, is pure isoscalar.
\item With one more step, from \eqref{VP56} we find $\Xi^-$:
\be
\Xi^-=\frac{1}{\sqrt{3}}\times \Big\{[ s_1^\phi,s_2^\phi]_1 d^\psi -\frac{1}{2}[d^\phi,s_1^\phi]_1 s_2^\psi -\frac{1}{2}[ d^\phi,s_2^\phi]_1 s_1^\psi -
\frac{\sqrt{3}}{2}[d^\phi,s_1^\phi]_0 s_2^\psi +\frac{\sqrt{3}}{2}[s_2^\phi,d^\phi]_0 s_1^\psi
~\Big\}
\label{Xi56}
\ee
which could also be obtained from \eqref{proton56} by the double change of name $u_1\to s_1,~u_2\to s_2$.
\end{itemize}


\begin{thebibliography}{99}

\bibitem{LHCb:2015yax}
R.~Aaij \textit{et al.} [LHCb],
Phys. Rev. Lett. \textbf{115} (2015), 072001

\bibitem{2019}
R.~Aaij \textit{et al.} [LHCb],
Phys. Rev. Lett. \textbf{122}, no.22, 222001 (2019)
doi:10.1103/PhysRevLett.122.222001
[arXiv:1904.03947 [hep-ex]].


\bibitem{m1} 
J.~J.~Wu, R.~Molina, E.~Oset and B.~S.~Zou,
Phys. Rev. Lett. \textbf{105}, 232001 (2010)
doi:10.1103/PhysRevLett.105.232001
[arXiv:1007.0573 [nucl-th]];
W.~L.~Wang, F.~Huang, Z.~Y.~Zhang and B.~S.~Zou,
Phys. Rev. C \textbf{84}, 015203 (2011)
doi:10.1103/PhysRevC.84.015203
[arXiv:1101.0453 [nucl-th]];
M.~Karliner and J.~L.~Rosner,
Phys. Rev. Lett. \textbf{115}, no.12, 122001 (2015)
doi:10.1103/PhysRevLett.115.122001
[arXiv:1506.06386 [hep-ph]];
R.~Chen, X.~Liu, X.~Q.~Li and S.~L.~Zhu,
Phys. Rev. Lett. \textbf{115}, no.13, 132002 (2015)
doi:10.1103/PhysRevLett.115.132002
[arXiv:1507.03704 [hep-ph]];
L.~Roca, J.~Nieves and E.~Oset,
Phys. Rev. D \textbf{92}, no.9, 094003 (2015)
doi:10.1103/PhysRevD.92.094003
[arXiv:1507.04249 [hep-ph]];
J.~He,
Phys. Lett. B \textbf{753}, 547-551 (2016)
doi:10.1016/j.physletb.2015.12.071
[arXiv:1507.05200 [hep-ph]].

{\it See also the reviews:} 
H.~X.~Chen, W.~Chen, X.~Liu and S.~L.~Zhu,
Phys. Rept. \textbf{639}, 1-121 (2016)
doi:10.1016/j.physrep.2016.05.004
[arXiv:1601.02092 [hep-ph]];
F.~K.~Guo, C.~Hanhart, U.~G.~Mei\ss{}ner, Q.~Wang, Q.~Zhao and B.~S.~Zou,
Rev. Mod. Phys. \textbf{90}, no.1, 015004 (2018)
[erratum: Rev. Mod. Phys. \textbf{94}, no.2, 029901 (2022)]
doi:10.1103/RevModPhys.90.015004
[arXiv:1705.00141 [hep-ph]];
Y.~R.~Liu, H.~X.~Chen, W.~Chen, X.~Liu and S.~L.~Zhu,
Prog. Part. Nucl. Phys. \textbf{107}, 237-320 (2019)
doi:10.1016/j.ppnp.2019.04.003
[arXiv:1903.11976 [hep-ph]].


\bibitem{m2} 
F.~K.~Guo, U.~G.~Mei\ss{}ner, W.~Wang and Z.~Yang,
Phys. Rev. D \textbf{92}, no.7, 071502 (2015)
doi:10.1103/PhysRevD.92.071502
[arXiv:1507.04950 [hep-ph]];
X.~H.~Liu, Q.~Wang and Q.~Zhao,
Phys. Lett. B \textbf{757}, 231-236 (2016)
doi:10.1016/j.physletb.2016.03.089
[arXiv:1507.05359 [hep-ph]];
M.~Mikhasenko,
[arXiv:1507.06552 [hep-ph]];
A.~P.~Szczepaniak,
Phys. Lett. B \textbf{757}, 61-64 (2016)
doi:10.1016/j.physletb.2016.03.064
[arXiv:1510.01789 [hep-ph]].


\bibitem{Maiani:2015vwa}
L.~Maiani, A.~D.~Polosa and V.~Riquer,
Phys. Lett. B \textbf{749} (2015), 289.

\bibitem{Ali}
A.~Ali, I.~Ahmed, M.~J.~Aslam and A.~Rehman,
Phys. Rev. D \textbf{94}, no.5, 054001 (2016)
doi:10.1103/PhysRevD.94.054001
[arXiv:1607.00987 [hep-ph]];
A.~Ali, I.~Ahmed, M.~J.~Aslam, A.~Y.~Parkhomenko and A.~Rehman,
JHEP \textbf{10}, 256 (2019)
doi:10.1007/JHEP10(2019)256
[arXiv:1907.06507 [hep-ph]];
A.~Ali, I.~Ahmed, M.~J.~Aslam, A.~Parkhomenko and A.~Rehman,
PoS \textbf{ICHEP2020}, 527 (2021)
doi:10.22323/1.390.0527
[arXiv:2012.07760 [hep-ph]].

  \bibitem{Lebed:2015tna}
R.~F.~Lebed,
Phys. Lett. B \textbf{749} (2015), 454-457;
[arXiv:1507.05867 [hep-ph]].

\bibitem{4q} 
G.~N.~Li, X.~G.~He and M.~He,
JHEP \textbf{12}, 128 (2015)
doi:10.1007/JHEP12(2015)128
[arXiv:1507.08252 [hep-ph]];
R.~Zhu and C.~F.~Qiao,
Phys. Lett. B \textbf{756}, 259-264 (2016)
doi:10.1016/j.physletb.2016.03.022
[arXiv:1510.08693 [hep-ph]].
For reviews 
A.~Esposito, A.~Pilloni and A.~D.~Polosa,
Phys. Rept. \textbf{668}, 1-97 (2017)
doi:10.1016/j.physrep.2016.11.002
[arXiv:1611.07920 [hep-ph]];
A.~Ali, J.~S.~Lange and S.~Stone,
Prog. Part. Nucl. Phys. \textbf{97}, 123-198 (2017)
doi:10.1016/j.ppnp.2017.08.003
[arXiv:1706.00610 [hep-ph]].

\bibitem{noiBO}L.~Maiani, A.~D.~Polosa and V.~Riquer,
  Phys.\ Rev.\ D {\bf 100} (2019)  074002.
   
\bibitem{weinbergQM} S. Weinberg, {\it  Lectures on Quantum Mechanics}, Cambridge University Press (2015).

\bibitem{pauling} L. Pauling, Chem. Rev., {\bf 5}, 173-213 (1928), DOI: 10.1021/cr60018a003, see also L. Pauling and E. B. Wilson Jr., {\it Introduction to Quantum Mechanics with Applications to Chemistry}, Dover Books on Physics (1985).  


 \bibitem{Bicudo:2012qt}
P.~Bicudo \textit{et al.} [European Twisted Mass],
Phys. Rev. D \textbf{87} (2013) no.11, 114511, arXiv:1209.6274. 






\bibitem{Gursey:1992dc}
F.~Gursey and L.~Radicati,
Phys. Rev. Lett. \textbf{13} (1964), 173; F.~Gursey, A.~Pais and L.~Radicati,
Phys. Rev. Lett. \textbf{13} (1964), 299.

   \bibitem{Santopinto:2016pkp}
E.~Santopinto and A.~Giachino,
Phys. Rev. D \textbf{96} (2017), 014014,
[arXiv:1604.03769 [hep-ph]].

  \bibitem{georgi} H. Georgi,  {\it Lie Algebras in Particle Physics}, Westviews Press (1999).


\bibitem{isgur} 
N.~Isgur and G.~Karl,
Phys. Rev. D \textbf{18}, 4187 (1978)
doi:10.1103/PhysRevD.18.4187.

 \bibitem{roberts} S. Capstick and W. Roberts, Prog. Part. Nucl. Phys. {\bf 45}, S241 (2000), [arXiv:nucl-th/0008028 [nucl-th]].
  

\bibitem{Bali:2000gf}
  G.~S.~Bali,
  Phys.\ Rept.\  {\bf 343} (2001) 1;
  [hep-ph/0001312]. 

\bibitem{Karliner:2013dqa}
M.~Karliner and S.~Nussinov,
JHEP \textbf{07} (2013), 153; [arXiv:1304.0345 [hep-ph]].


\bibitem{Eichten:2017ffp}
E.~J.~Eichten and C.~Quigg,
Phys. Rev. Lett. \textbf{119} (2017) 202002; [arXiv:1707.09575 [hep-ph]].

\bibitem{Maiani:2022qze}
L.~Maiani, A.~Pilloni, A.~D.~Polosa and V.~Riquer,
Phys. Lett. B \textbf{836} (2023), 137624; arXiv:2208.02730 [hep-ph].



 \end{thebibliography}
 \end{document}